\documentclass[aps,prd,twocolumn,superscriptaddress]{revtex4-1}
\usepackage[margin=2.0cm]{geometry}
\usepackage{amsmath, amsthm, amssymb, bm}
\usepackage{mathrsfs}
\usepackage{graphics}
\usepackage{graphicx}
\usepackage{color}
\usepackage{indentfirst}
\usepackage{natbib}
\usepackage{hyperref}
\begin{document}

\title{Radiative Seesaw Model with Non-zero $ \theta_{13} $ and Warm Dark Matter Scenario}
\author{Ping-Kai Hu}
\affiliation{Department of Physics, National Tsing Hua University, HsinChu 300, Taiwan}
\date{\today}
\begin{abstract}
With updated experiment result, we study Ma's radiative seesaw model.
We try to use fewer assumptions to study the feasibility of this model.
Both CDM and WDM scenario are taken into consideration.
In CDM scenario, 
the fitting of relic abundance shows that a mass spectrum $ M_1 < m_0 < M_2 $ is preferred if theory remains perturbability.
By studying the flavour violation process $ \mu \rightarrow e \gamma $ and relic density abundance,
we get very strong constraint for the model.
In addition, we try to explain muon anomaly $ a_{\mu} \equiv (g-2)/2 $ within the model, 
but the contribution induced by new particles is too small to account for the discrepancy.
In WDM scenario,
We show that relic abundance can be produced by thermal production with subsequent entropy dilution process.
The entropy dilution can be attained by $ N_2 $ decay in which a mass spectrum $ M_1 < M_2 < m_0 $ is required.
\end{abstract}
\maketitle
%
%
\section{Introduction}

In physics and cosmology, dark matter and neutrino mass are two outstanding problems for Standard Model.
Many evidences show that our universe contains a great amount of non-luminous matter \cite{Komatsu:2010fb}. 
The observation of neutrino oscillation implies non-zero neutrino masses \cite{Fukuda:1998mi,*Ahmad:2001an,*Ahmad:2002jz}.
These results are most important evidences indicating the physics beyond Standard Model.

Small neutrino mass can be generated in several ways.
Certainly seesaw mechanism is the most attractive one for generating neutrino mass \cite{Minkowski:1977sc,*yanagida:1979, *gell-mann:1979, *Cheng:1980qt, *Schechter:1980gr}.
So far we still do not know the nature of dark matter.
However, we can realize some properties of dark matter from observation, 
such as its gravitational effects and that it must be weakly interacting.
One essential property is that it must be stable.
To ensure the stability of dark matter, 
introducing a $ Z_2 $ symmetry is a simple approach which is employed in many models.
In this scheme, the lightest $ Z_2 $-odd particle cannot decay into other particles, serving as dark matter candidate.
The best-known example is the $ R $-parity in supersymmetric models. 
Since both neutrino mass and dark matter problem are two very important issues while exploring beyond Standard Model regime,
we want to investigate the possibility that they might be related.
This leads to many models trying to solve these problems simultaneously.

One of these is the radiative seesaw model proposed by Ma \cite{Ma:2006}
which gives a simple way to explain both the dark matter problem and smallness of neutrino mass. 
It extends Standard Model with a $ Z_2 $ symmetry, 
and uses one loop effect to generate small neutrino masses.
In addition, one scalar doublet and three Majorana fermions are introduced.
So this model is a modest extension for getting dark matter and neutrino mass.

Some works have already been done along this line. 
Sierra \textit{et al} \cite{Sierra:2009} 
consider a specific mass spectrum, and discuss the dark matter phenomena.
They deduce some collider signatures which can be tested in LHC.  
They also discuss warm dark matter (WDM) scenario, but fail to explain the relic abundance.
Suematsu \textit{et al} \cite{Suematsu:2009ww} 
discuss cold dark matter (CDM) abundance by studying the structure of mass mixing matrix and coannihilation effect.
In addition, they take into account the $ \mu \rightarrow e \gamma $ process to get constraint. 
However, 
their analysis assumes that $ \theta_{13} = 0 $ 
which has been found to be non-zero in the recent Daya-bay's result. 
In this paper, we want to re-examine Ma's model in light of this new experimental result.

Recently, many experiments on neutrino physics have great progress.
A large amount of data has been accumulated 
so that we are able to determine mixing angle of mass matrix and mass difference more precisely. 
In last two years,  
experiments including T2K \cite{T2K:2011qn}, Daya-bay \cite{DAYA:2012eh}, and RENO \cite{RENO:2012nd}
present new data about $ \theta_{13} $.
These results confirm non-zero $ \theta_{13} $ and one CP-violation phase needs to be studied.
With these information,
we can now use neutrino mass mixing matrix to study Ma's radiative seesaw model more closely.

Recently, WDM scenario gains considerable 
attention \cite{Bezrukov:2009th, King:2012wg, Robinson:2012wu}. 
Some literatures argue that WDM scenario might be able to overcome some problems 
that typical CDM have met in structure formation and galaxy halos
\cite{Moore:1999gc, Bode:2000gq, AvilaReese:2000hg}.
As a result, we shall take WDM scenario into consideration as well.
WDM can be attained by several different mechanisms. 
For example,
Dodelson and Widrow \cite{Dodelson:1993je} 
propose that WDM can be produced by active-sterile neutrino oscillation during big bang nucleosynthesis (BBN) epoch.  
Shi and Fuller \cite{Shi:1998km} propose that WDM can be accomplished by a non-thermal resonant production. 
In this model, there is no mixing with active neutrino. 
Hence, Dodelson and Widow's approach cannot apply to this case.  
Since the Yukawa coupling of the sterile neutrinos in this model is not necessarily small, 
we can realize the WDM scenario of Ma's model by thermal production \cite{Scherrer:1984fd}.
Some researches show that correct dark matter relic density can be obtained by thermal overproduction with a subsequent entropy dilution from some decay channel \cite{King:2012wg}.

In Ref. \cite{Ma:2012if}, the WDM of radiative seesaw model from thermal production has been studied.
However, its entropy dilution is accomplished by adding another real scalar singlet which couples to Higgs doublet. 
Actually extra particle for providing reheating process is not really needed.
We try to realize required entropy dilution within this model, without any additional particle.

In this paper, we focus on the Ma's radiative seesaw model, 
trying to study its phenomena.
With updated experimental result, we can reconstruct the mass matrix.
Different from previous works, we use another method to analyze mass mixing matrix.
Both CDM and WDM would be taken into account. 
By studying the flavour violation process, we would put constraint on the parameter space of this model.
In addition, we pay attention to muon anomaly and try to explain the experimental result within the model.

This paper is arranged as follows.
In next Section, we briefly introduce the radiative seesaw model.
In Section III, we study the relic abundance.
In Section IV and V, we study the phenomena of flavour violation process and muon anomaly.
Finally we give a summary in Section VI.

\section{Model}
Ernest Ma's radiative seesaw model \cite{Ma:2006} extends Standard Model symmetry into : 
$ SU(2) \times U(1) \times Z_2 $
where a discrete $ Z_2 $ symmetry is introduced.
In addition, one scalar $ SU(2) $ doublet $ \eta = ( \eta^+ , \eta ^0 ) $ and three Majorana fermion $ N_i $ ($ i = 1, 2, 3 $) are added. 
All the new particles are assigned as $ Z_2 $-odd.
Therefore, both neutral particle $ \eta^0 $ and the lightest Majorana fermion $ N_1 $ could serve as the dark matter candidate. 
Here we suppose the dark matter is $ N_1 $.
With these additional particles, the Yukawa coupling of leptons and scalar sector have to be modified. 
The Yukawa coupling of leptons is given by
\begin{equation}
\mathcal{L}_{Y} 
= f_{\alpha \beta} (\bar{\nu}_{\alpha} \phi^{+} + \bar{l}_{\alpha} \phi^0) l_{\beta}
+ h_{\alpha i}     (\bar{\nu}_{\alpha} \eta^0 - \bar{l}_{\alpha} \eta^{+}) N_i 
+ \mathrm{h.c.}
\end{equation} 
where $ \alpha , \beta = e, \mu, \tau $, and 
$ h_{\alpha i} $ are unknown parameters. 
Majorana mass term of $ N_i $ is given by
\begin{equation}
\frac{1}{2} M_i \bar{N}^c_i N_i + \mathrm{h.c.}
\end{equation}
where $ M_1 < M_2 < M_3 $. 
\\
\indent
The Higgs potential of this model is given by
\begin{align}
V (\Phi , \eta) =
&\mu^2 \Phi^{\dagger} \Phi + \mu_2^2 \eta^{\dagger} \eta
+ \lambda_1 (\Phi^{\dagger} \Phi)^2
+ \lambda_2 (\eta^{\dagger} \eta)^2
\notag 
\\
&+ \lambda_3 (\Phi^{\dagger} \Phi)(\eta^{\dagger} \eta) 
+ \lambda_4 (\Phi^{\dagger} \eta) (\eta^{\dagger} \Phi)
\notag
\\
&+ \frac{1}{2} \lambda_5 [(\Phi^{\dagger} \eta)^2 + \mathrm{h.c.}] 
\end{align}
where $ \lambda_5 $ is taken to be real without losing any generality.
In this potential, we will take $ \mu_2^2 > 0 $ 
so that only $ \phi^0 $ get a non-zero vacuum expectation value $ \langle \phi^0 \rangle = v $ 
which causes spontaneous symmetry breaking(SSB)
and $ \eta^0 $ will not develop vacuum expectation value to keep the $ Z_2 $ symmetry unbroken.
The neutral component of $ \eta $ can be separated into real part and imaginary part, 
i.e. $ \eta^0 = ( \eta^0_R + i \eta^0_I ) / \sqrt{2} $.
From SSB, the masses of physical scalar bosons would be split, 
and the values are given by
\begin{align}
m^2( \eta^{\pm} ) &= \mu_2^2 + \lambda_3 v^2 
\notag \\
m^2( \eta_{\mathrm{R}}^0 ) &= \mu_2^2 + \lambda_3 v^2 +(\lambda_4 + \lambda_5) v^2
\\
m^2( \eta_{\mathrm{I}}^0 ) &= \mu_2^2 + \lambda_3 v^2 +(\lambda_4 - \lambda_5) v^2.
\notag
\end{align}
Since $ \lambda_5 $ is related to the active neutrino masses, it' s assumed to be very small. 
( The relationship between $ \lambda_5 $ and neutrino masses will be discussed later. )
Hence, for simplicity, 
we assume the masses of $ \eta^0_R $ and $ \eta^0_I $ to be degenerate, 
and their mass is given by $ m_0^2 = \mu_2^2 + (\lambda_3 + \lambda_4) v^2 $.
\begin{figure}[ht]
\begin{center}
\includegraphics[width=0.6\linewidth]{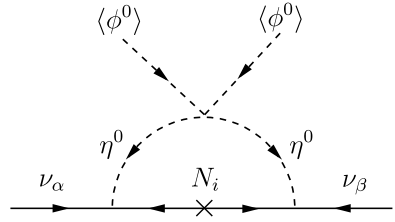}
\end{center}
\caption{
One-loop generation of Majorana neutrino mass.
}
\label{mass_generation}
\end{figure}

Due to $ Z_2 $ symmetry, Dirac mass of neutrinos is forbidden. 
Neutrino masses can only be generated by one-loop radiative effect involving $ \eta^0 $ and $ N_i $
as showed in FIG. \ref{mass_generation}.
And the generated mass mixing matrix of neutrino is given by \cite{Suematsu:2009ww} 
\begin{equation}
\mathcal{M}_{\alpha \beta} = \sum^{3}_{i=1} h_{\alpha i} h_{\beta i} \Lambda_i
\label{mass_mixing}
\end{equation}
where $ \Lambda_i $ is given by
\begin{equation}
\Lambda_i = \dfrac{\lambda_5 v^2}{8 \pi^2 M_i} I \left( \dfrac{M_i}{m_0} \right) ,
\end{equation}
\begin{equation}
I(x) = \dfrac{x^2}{1-x^2} \left( 1 + \dfrac{x^2}{1-x^2} \mathrm{ln} x^2 \right).
\end{equation}

Here we see that we can take $\lambda _{5}$ to be small to get small neutrinos masses.
In this model,
there are nine Yukawa coupling constants, $ h_{\alpha i} $.
In order to make our analysis simpler, 
we will reduce the number of unknown parameters. 
Since recent experiments of neutrino oscillation measure the mixing angles and mass differences more accurately, 
we can use these results to fix some of our parameters.

In general, this mass mixing matrix in eq. (\ref{mass_mixing}) can be diagonalized by a 
Pontecorvo-Maki-Nakagawa-Sakata Matrix (PMNS) 
which is of the form 
\begin{widetext}
\begin{equation}
U =
\left(
\begin{array}{ccc}
c_{12}c_{13} & s_{12}c_{13} & s_{13}e^{-i\delta} \\
-s_{12}c_{23}-c_{12}s_{23}s_{13}e^{i\delta} & c_{12}c_{23}-s_{12}s_{23}s_{13}e^{i\delta} & s_{23}c_{13} \\
-s_{12}s_{23}-c_{12}c_{23}s_{13}e^{i\delta} & -c_{12}s_{23}-s_{12}c_{23}s_{13}e^{i\delta} & c_{23}c_{13}
\end{array}
\right)
\end{equation}
\end{widetext}
where $ c_{ij} = \mathrm{cos} \theta_{ij}, s_{ij} = \mathrm{sin} \theta_{ij} $ and
$ \delta $ is the CP violation phase. 
The diagonal mass matrix is given by
\begin{equation}
U^{T} \mathcal{M} U = \mathrm{diag}(m_1,m_2,m_3)
\end{equation}
where $ m_1 $, $ m_2 $, and $ m_3 $ are the masses of neutrinos. 
In general, there could be two more Majorana phases in this formula.
Nevertheless, in order to simplify our analysis, we take them to be zero.

According to the most updated fit \cite{Fogli:2012ua}, the mixing angles are
$ \mathrm{sin}^2 \theta_{12} = 3.07 \times 10^{-1} $, 
$ \mathrm{sin}^2 \theta_{23} = 3.98 \times 10^{-1} $, and
$ \mathrm{sin}^2 \theta_{13} = 2.45 \times 10^{-2} $.
The CP violation phase $ \delta = 0.89 \pi $.
The mass square differences are
$ \Delta m_{21}^2 = 7.54 \times 10^{-5} (\mathrm{eV})^2 $ and 
$ \Delta m_{31}^2 = 2.43 \times 10^{-3} (\mathrm{eV})^2 $.
Here the normal neutrino mass hierarchy is assumed, i.e. $ m_1 < m_2 < m_3 $.
From the cosmological observation of WMAP, there is an upper bound for total neutrino mass, 
$ \sum m_{\nu} < 0.58 \: \mathrm{eV} $ (95\% CL) \cite{Komatsu:2010fb}.
Here we assume the extreme situation: 
$ m_1 = 0 $, 
$ m_2 = 8.68 \times 10^{-3} $ eV, and 
$ m_3 = 0.049 $ eV.
By using these data and some assumptions, we can construct the mass mixing matrix as
\begin{align}
\mathcal{M} &\cong
\left(
\begin{array}{ccc}
1.18+0.37i & 0.96+0.16i & 2.58+0.50i 
\\
0.96+0.16i & 1.24+0.005 & 1.40+0.14i 
\\
2.58+0.50i & 1.40+0.14i & 6.46+0.41i
\end{array}
\right)
\notag
\\
&\qquad \qquad \qquad \qquad \qquad \qquad \qquad \quad \times 10^{-2} \mathrm{eV}.
\end{align}
As we know, the value of $ \Lambda_i $ depends on the ratio, $ (M_i / m_0) $. 
If $ M_3 $ is much larger than $ M_1 $, $ M_2 $ and $ m_0 $, 
$ \Lambda_3 $ would be negligible compared to $ \Lambda_1 $ and $ \Lambda_2 $. 
It means that we can neglect the contribution from $ \Lambda_3 $ and $ h_{\alpha 3} $ in this stage.
In this way, we reduce the number of parameters that we have to handle.

We define a real positive number $ \alpha $,
\begin{equation}
\alpha \equiv \dfrac{\Lambda_2}{\Lambda_1}
\end{equation}
and five complex parameters,
\begin{equation}
a_1 \equiv \dfrac{h_{\mu 1}}{h_{e 1}} , \: 
a_2 \equiv \dfrac{h_{\tau 1}}{h_{e 1}} 
\end{equation}
\begin{equation}
b_1 \equiv \dfrac{h_{e 2}}{h_{e 1}} , \: 
b_2 \equiv \dfrac{h_{\mu 2}}{h_{e 1}} , \: 
b_3 \equiv \dfrac{h_{\tau 2}}{h_{e 1}} 
\end{equation}
By using the values of mixing matrix, we can get
$ a_1 \cong 0.70 - 0.72 i $ and 
$ a_2 \cong 2.27 + 0.61 i $, 
which are useful in dealing with the relic density. 
$ b_1 $, $ b_2 $, and $ b_3 $ cannot be exactly solved, depending on $ \alpha $.
Since the mixings in neutrino mass matrix are large,
it's reasonable to take coupling constants to be of roughly of same order of magnitude.
Although we take the extreme situation that $ m_1 = 0 $, 
we can also consider other situations and get similar results.

\section{Relic abundance}
%
%
\subsection{Cold dark matter scenario}
In order to explain the dark matter problem, 
it's necessary for a model to have right amount of relic abundance. 
Here, we first discuss the CDM scenario, 
and we will also study the WDM in the next subsection. 
At early universe, the temperature is very high.
Various particles are produced and annihilate, reaching equilibrium quickly.
As universe expands, the temperature would drop, 
and particles would decouple from each other.
Since $ N_1 $ is absolutely stable, the abundance from thermal production would remain as dark matter relic.
To discuss the magnitude of density, 
we need to study the Boltzmann equation given by \cite{Kolb:1990vq}
\begin{equation}
\dfrac{dn}{dt} + (3H)n = 
- \langle \sigma \left| \vec{v} \right| \rangle
(n^2 - n_{\mathrm{eq}}^2)
\end{equation}
where $ n $ is the number density of dark matter, 
$ H $ is the Hubble constant, 
and $ \sigma $ is the annihilation cross-section of dark matter particles.
$ \vec{v} $ indicates the relative velocity of two annihilation particles.
$ n_{\mathrm{eq}} $ means the number density in equilibrium which yields to Boltzmann distribution.

In CDM scenario, dark matter particles are non-relativistic at decoupling.
In general, the annihilation cross-section takes on this form: 
$ \langle \sigma \left| \vec{v} \right| \rangle \cong a_{\mathrm{eff}} + b_{\mathrm{eff}} \langle \left| \vec{v} \right|^2 \rangle $ 
where terms of higher order in $ \left| \vec{v} \right| $ are neglected.
The approximate solution for Boltzmann equation is given by \cite{Kolb:1990vq}
\begin{equation}
\Omega_{\mathrm{DM}} = \sqrt{\dfrac{45}{\pi }} g_*^{-1/2} 
\dfrac{x_{\mathrm{F}} S}{ M_{\mathrm{pl}} (a_{\mathrm{eff}} + 3 b_{\mathrm{eff}}/ x_{\mathrm{F}} ) \rho_c }
\: \:
,\: x_{\mathrm{F}} \equiv \dfrac{M_1}{T_{\mathrm{F}}}
\end{equation}
where $ T_{\mathrm{F}} $ is freeze-out temperature at decoupling,
$ g_* $ is the number of degrees of freedom(DOF) at freezing-out, 
$ S $ is entropy density,
$ M_{\mathrm{pl}} $ is Plank mass, 
and $ \rho_c $ is critical energy density of the universe,.

In the nonrelativistic limit, 
the annihilation cross-section of $ N_1 N_1 \rightarrow l_{\alpha} l_{\beta} $ multiplied by the relative velocity is given by
\cite{Suematsu:2009ww}
\begin{equation}
\sigma \left| \vec{v} \right| =
\dfrac{1}{48 \pi} \dfrac{M_1^2 (M_1^4 + m_0^4)}{ (M_1^2 + m_0^2)^4 } \left| \vec{v} \right|^2 
\sum_{\alpha \beta} \left( h_{\alpha 1}^2 h^{*2}_{\beta 1} + \mathrm{h.c.} \right)
\end{equation}
\begin{figure}[h]
\begin{center}
\includegraphics[angle=-0,width=0.8\linewidth]{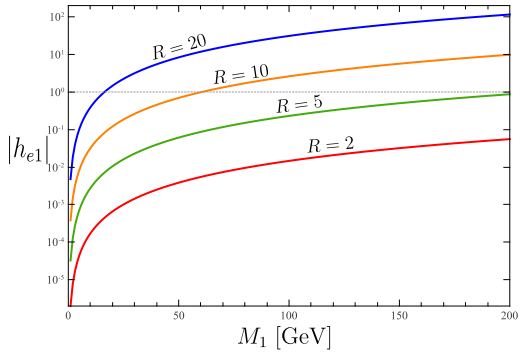}
\caption{
Relationship $ |h_{e 1}| - M_1 $ where $ R = m_0 / M_1 $. 
This plot is based on the non-baryon relic density 
$ \Omega_{\mathrm{DM}} h^2 = 0.110 $.
Different values of $ R = $ 2, 5, 10, 20 are considered. 
} 
\label{h_to_M}
\end{center}
\end{figure}
\\
There are two Feynman diagrams(the lowest order) which contribute to this cross-section.
Because of the Fermi statistics there is a minus sign between these diagrams and 
as a result the s-wave parts of the scattering cancel out. 
Only p-wave part remains. 
With the value of cross-section, 
we can get the relationship between $ |h_{e1}| $ and $ M_1 $. 
Using current data from several observations,
the non-baryon relic density   is 
$ \Omega_{\mathrm{DM}} h^2 = 0.110 \pm 0.006 $ with 68 \% C.L. \cite{Lahav:2010mi}.
Fitting the relic abundance, we get the fit in FIG. \ref{h_to_M}.
In FIG. \ref{h_to_M} we plot the relation between $ |h_{e1}| $ and $ M_1 $ for different values of $ R \equiv m_0/M_1 $.
Cases $ R = $ 2, 5, 10, 20 are considered. 
In some region of $ M_1 $, the required value of $ |h_{e1}| $ is larger than 1. 
It means that we have to deal with a non-perturbative interaction. 
This kind of situation is even worse for larger $ R $ cases.
We can see that in $ R = 10 $ case, the coupling is beyond perturbative regime for $ M_1 > $ 60 GeV.
In $ R = 20 $ case, the coupling is beyond perturbative regime for $ M_1 > $ 20 GeV, 
and our calculation of cross-section is probably not reliable. 
Nevertheless, we can conclude that a coupling larger than 1 is needed in these regions.
If we suppose the Yukawa interaction is perturbable,
a scenario $ R \lesssim 10 $ is preferred.
Since there is no reason to require $ M_1 $ and $ M_2 $ to be of the same order, 
$ M_2 > 10 M_1 $ is an acceptable choice.
Hence, a mass spectrum $ M_1 < m_0 < M_2 $ is preferred.

In following sections, we will take some phenomena into consideration, 
and constrain the relationship we get from relic density fitting.

\subsection{Warm dark matter scenario}
We now study the WDM scenario. 
where the dark matter particle has a mass of few keV.
Since the mass is quite small, the particle would still be relativistic at freeze-out point.  
The quantity $ Y \equiv n / S $ indicating the ratio between number density $ n $ and entropy density $ S $ is given by \cite{Kolb:1990vq}
\begin{equation}
Y_{N_1} = \dfrac{45 \zeta (3)}{2 \pi^4} \dfrac{ g_{\mathrm{eff}} }{g_{*s}}
\end{equation}
where $ g_{\mathrm{eff}} $ is the effective number of DOF of $ N_1 $ and $ g_{*s} \sim 10^2 $ is the the number of DOF for the entropy.  
To get the correct non-baryon abundance, $ Y_{N_1} $ must satisfy the condition 
$ Y_{N_1} M_1 S = \Omega_{\mathrm{DM}} \rho_{c} $ 
where $ \rho_{c} $ is the critical density and $ \Omega_{\mathrm{DM}} $ is the density fraction of non-baryon dark matter. 
The required value for $ g_{\mathrm{eff}} $ is given by
\begin{equation}
g_{\mathrm{eff}} = 0.144 \left( \dfrac{1 \mathrm{keV}}{M_1} \right).
\end{equation}
For sterile neutrinos, there exists a lower bound around 5.6 keV from high-redshift Lyman-$ \alpha $ forest observations \cite{Viel:2007mv}.
If $ M_1 \sim 10 $ keV, $ g_{\mathrm{eff}} $ is about 2-order larger than the required value. 
Thus, this value is too large for correctly accounting for current relic abundance.
Fortunately, It's possible to use an out-of-equilibrium decay to get around the problem. 
In this case, we can have other particle which has been decoupled and
would decay into other relativistic particles, thermalizing the universe.
Under the decay process, extra entropy would be produced. 
Therefore, during the process, entropy density would increase:
\begin{equation}
S \rightarrow \gamma S \: , \: Y \rightarrow \dfrac{Y}{\gamma} 
\end{equation}
where $ \gamma $ is the factor which labels the entropy change.
And the value of $ Y $ also changes under the decay process, decreasing by a factor.

If $ M_1 < m_0 < M_2 $, 
the major decay channel of $ N_2 $ would be $ N_2 \rightarrow \eta^0 \nu $ and $ N_2 \rightarrow \eta^+ l^- $. 
The decay widths are given by
\begin{align}
\Gamma ( N_2 \rightarrow \eta^0 \nu ) 
&=
\Gamma ( N_2 \rightarrow \eta^{\pm} l^{\mp} )
&\cong \dfrac{M_2}{8 \pi} \sum_{\alpha = e, \mu , \tau } \left| h_{\alpha 2} \right|^2
\end{align}
The scalar particles are also unstable, decaying into $ N_1 $. 
The decay widths are given by
\begin{equation}
\Gamma ( \eta^0 \rightarrow N_1 \nu ) 
=
\Gamma ( \eta^{\pm} \rightarrow N_1 l^{\pm} )
\cong \dfrac{m_0}{8 \pi} \sum_{\alpha = e, \mu , \tau } \left| h_{\alpha 1} \right|^2
\end{equation}
With these decay rates, following Ref. \cite{Scherrer:1984fd}, we can estimate the value of $ \gamma $,
\begin{equation}
\gamma \cong 1.83 \, \, g_*^{1/4} \dfrac{ M_2 Y_{N_2} }{ \sqrt{\Gamma M_{pl}} }
\end{equation}
where $ \Gamma $ is the total decay width of $ N_2 $.
For $ M_2 \sim $ 1 GeV, the Yukawa coupling constant is $ h_{\alpha i} \sim 10^{-12}$. 
However, since $ N_i $'s have no mixing with active neutrinos,
WDM may not be attained by thermal production with such small couplings \cite{Dodelson:1993je}.
Moreover, if we have such small couplings, 
we could have neutrino with around 1 eV Dirac mass in SM scenario by adding a right-handed Dirac neutrino. 
Instead of $ N_2 $, $ \eta $ decay is also a candidate for serving the reheating process.
But the decay of $ \eta $ has been discussed before \cite{Gelmini:2009xd}, and it has the same problem. 
A very small Yukawa coupling $ < 10^{-9} $ is needed for correct relic density.
As a result, it motivates us to consider other scenarios.
For $ M_1 < M_2 < m_0 $,
$ N_2 $ decays into $ N_1 $ and two leptons via an virtual $ \eta $.
We can estimate the decay width of this process
\begin{align}
\Gamma ( N_2 \rightarrow N_1 l \bar{l} ) 
&= \sum_{\alpha , \beta} \Gamma ( N_2 \rightarrow N_1 l_{\alpha}^+ l_{\beta}^- ) + \Gamma ( N_2 \rightarrow N_1 \nu_{\alpha} \bar{\nu_{\beta}} ) 
\notag \\
&\cong \sum_{\alpha, \beta} \left| h_{2 \alpha} h_{1 \beta} \right|^2 \dfrac{ M_2^5}{1122 \pi^3 m_0^4} 
\end{align}
\begin{figure}[t]
\begin{minipage}[t]{\linewidth}
\begin{center}
\includegraphics[width=0.8\linewidth]{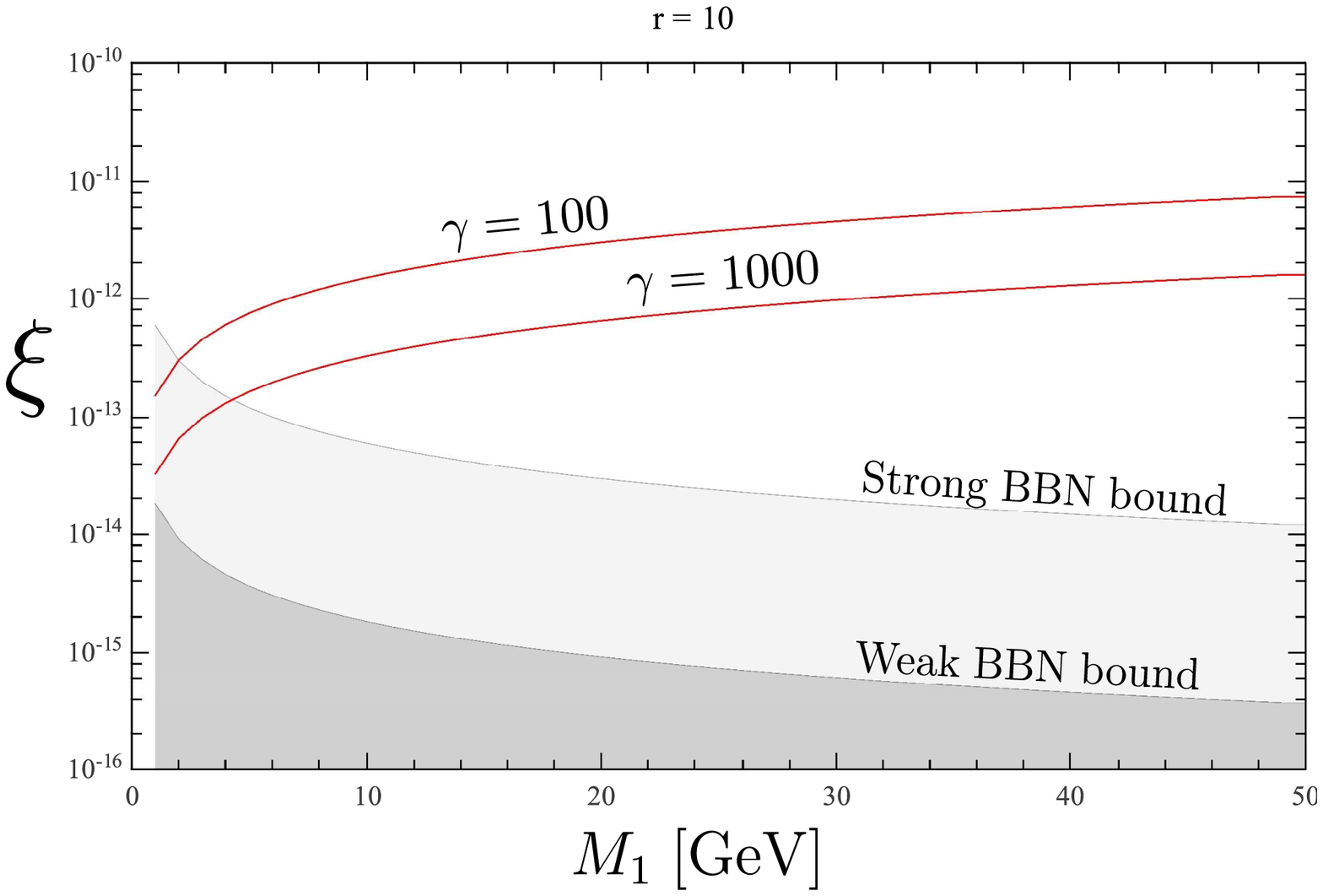}
\label{WDM10}
\end{center}
\end{minipage}
\begin{minipage}[t]{\linewidth}
\begin{center}
\includegraphics[width=0.8\linewidth]{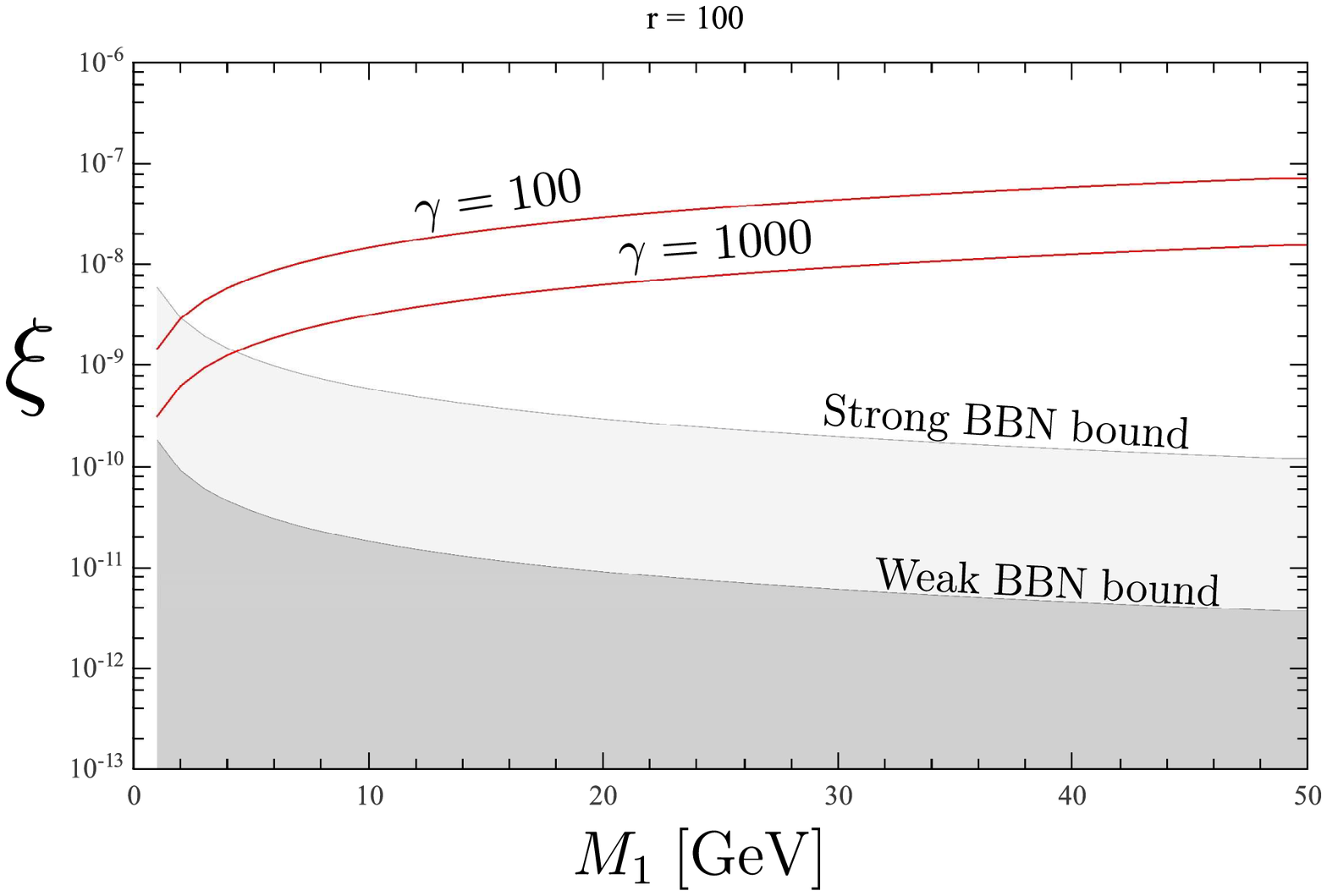}
\label{WDM100}
\end{center}
\end{minipage}
\caption{
Relationship between $ \xi $ and $ M_2 $ with strong and weak BBN bound.
Different values of $ r \equiv m_0 / M_2 $ are taken into account.
We consider the cases of $ \gamma = 100 $ and $ \gamma = 1000 $.
}
\label{WDM}
\end{figure}
where the mass of $ \eta^{\pm} $ are assumed to be $ m_0 $ for convenience.
This decay channel shows great potential to dilute the relic density and give the correct value of $ Y_{N_1} $. 
Here, one essential constraint from BBN has to be taken into account.
In order to preserve the normal BBN, the lifetime of $ N_2 $ should not be too long. 
In this way, the decay process could end before nucleons freeze out.
We can estimate the reheating temperature to be \cite{Scherrer:1984fd}, 
\begin{equation}
T_{\mathrm{rh}} \cong \dfrac{1}{2} \left( \dfrac{45}{2\pi g_*} \right)^{1/4} \sqrt{\Gamma M_{pl}}
\end{equation}
where $ \Gamma $ is the decay width of processes $ N_2 \rightarrow N_1 l \bar{l} $.
$ T_{\mathrm{rh}} $ must be larger than around 
0.7 MeV \cite{Kawasaki:2000en} (weak BBN bound)
to
4.0 MeV \cite{Hannestad:2004px} (strong BBN bound).
The relationship between coupling constant $ \xi \equiv \sum \left| h_{2 \alpha} h_{1 \beta} \right|^2 $ and $ M_2 $ is plotted in FIG. \ref{WDM} with the lower bounds. 
Here $ r $ indicates the ratio ($ m_0 / M_2 $), and different values of $ r $ are considered in the plots.   
From the plots, we can see that only a small region has been excluded. 
Except for $ M_2 $ being around a few GeV, most area of $ M_2 $ is safe for both weak and strong BBN bounds. 
It shows that considering $ N_1 $ as a WDM candidate is a viable scenario. 

\section{Lepton flavour violation}
\begin{figure}[h]
\includegraphics[width=0.6\linewidth]{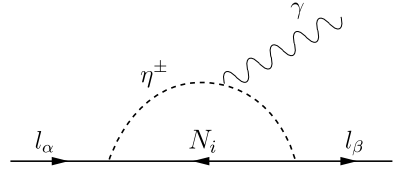}
\caption{
Flavour violation process induced by $ \eta $ and $ N_i $.
}
\label{flavour_violation}
\end{figure}

\begin{figure}[t]
\begin{center}
\includegraphics[width=0.8\linewidth]{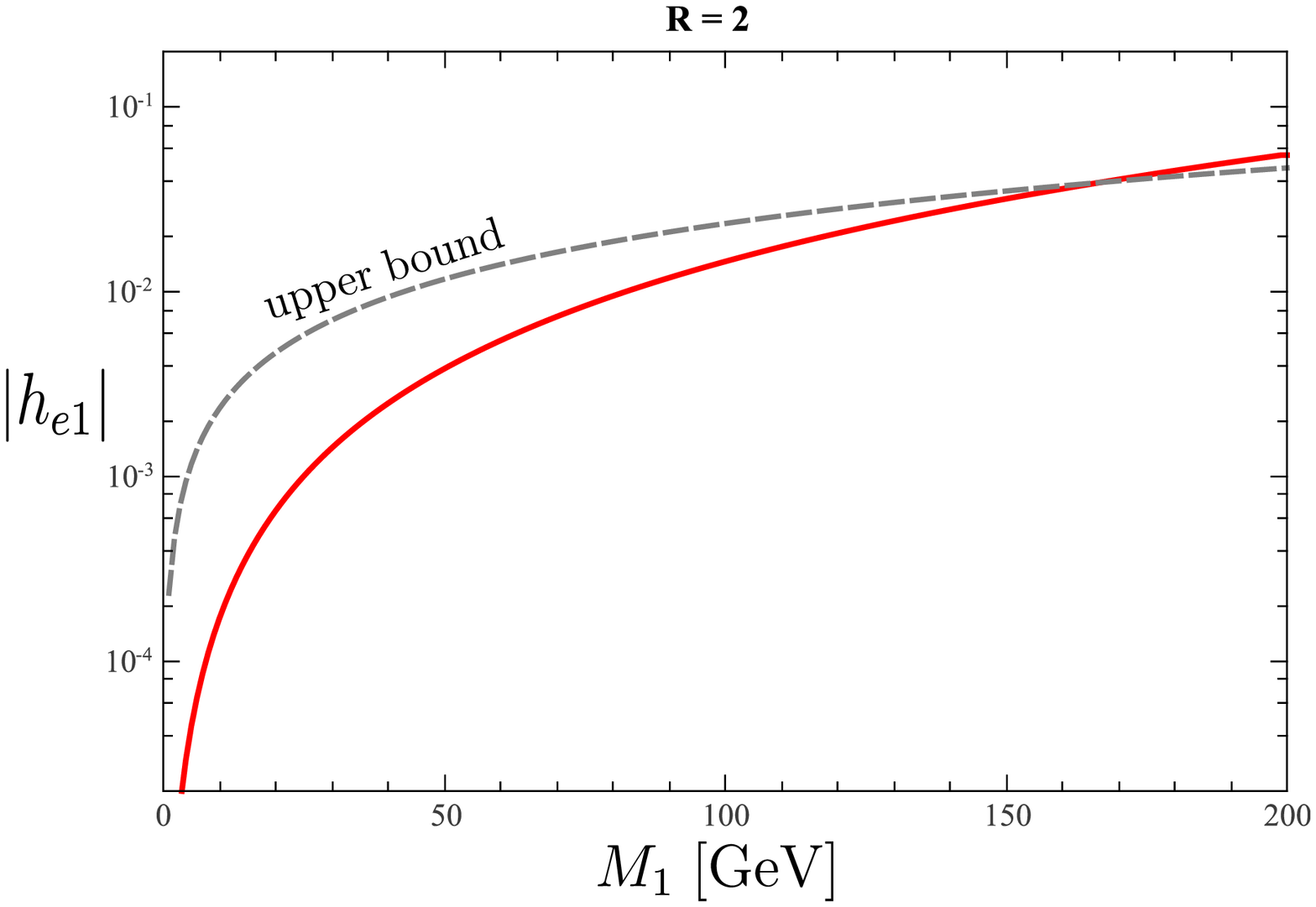}
\includegraphics[width=0.8\linewidth]{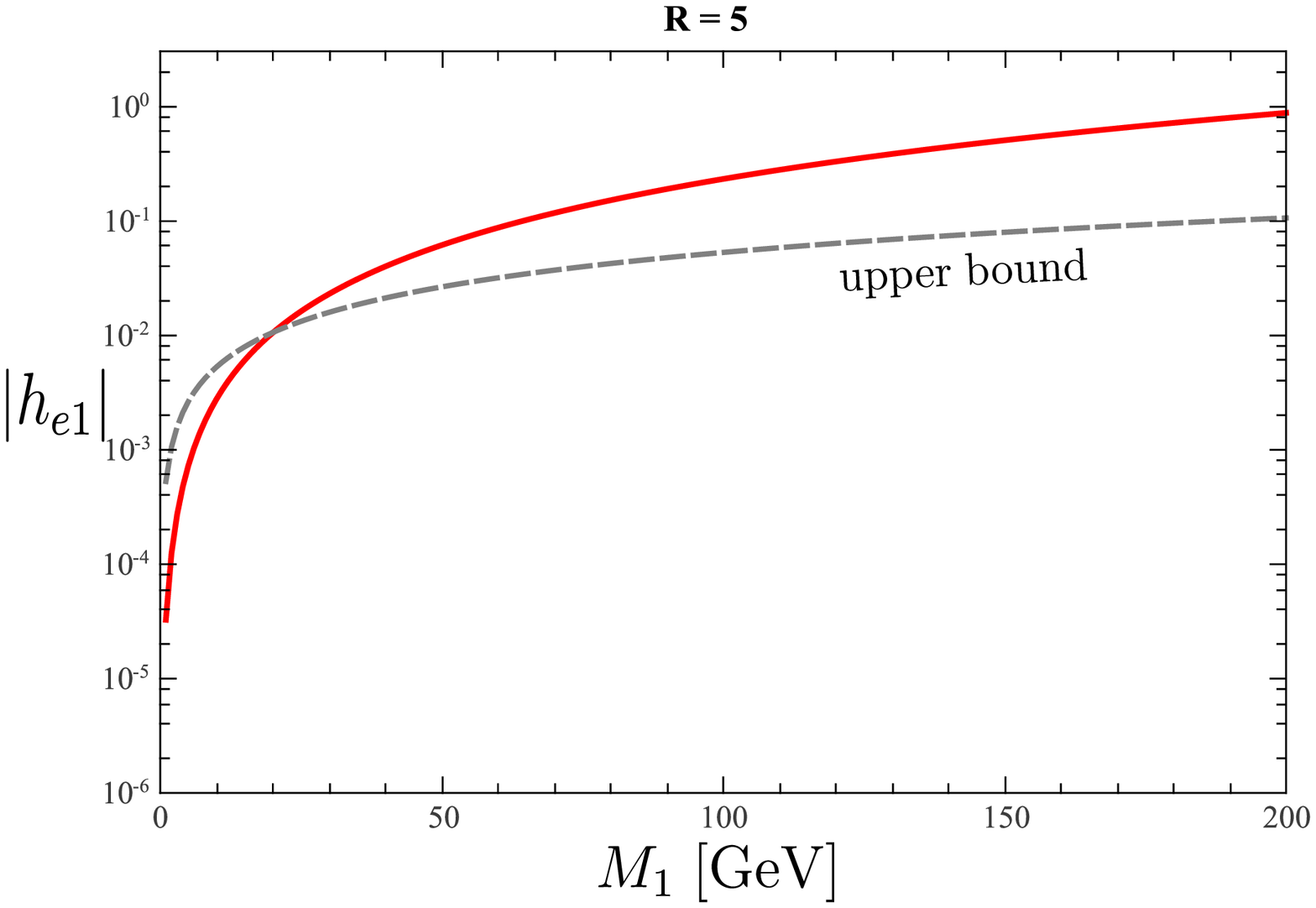}
\includegraphics[width=0.8\linewidth]{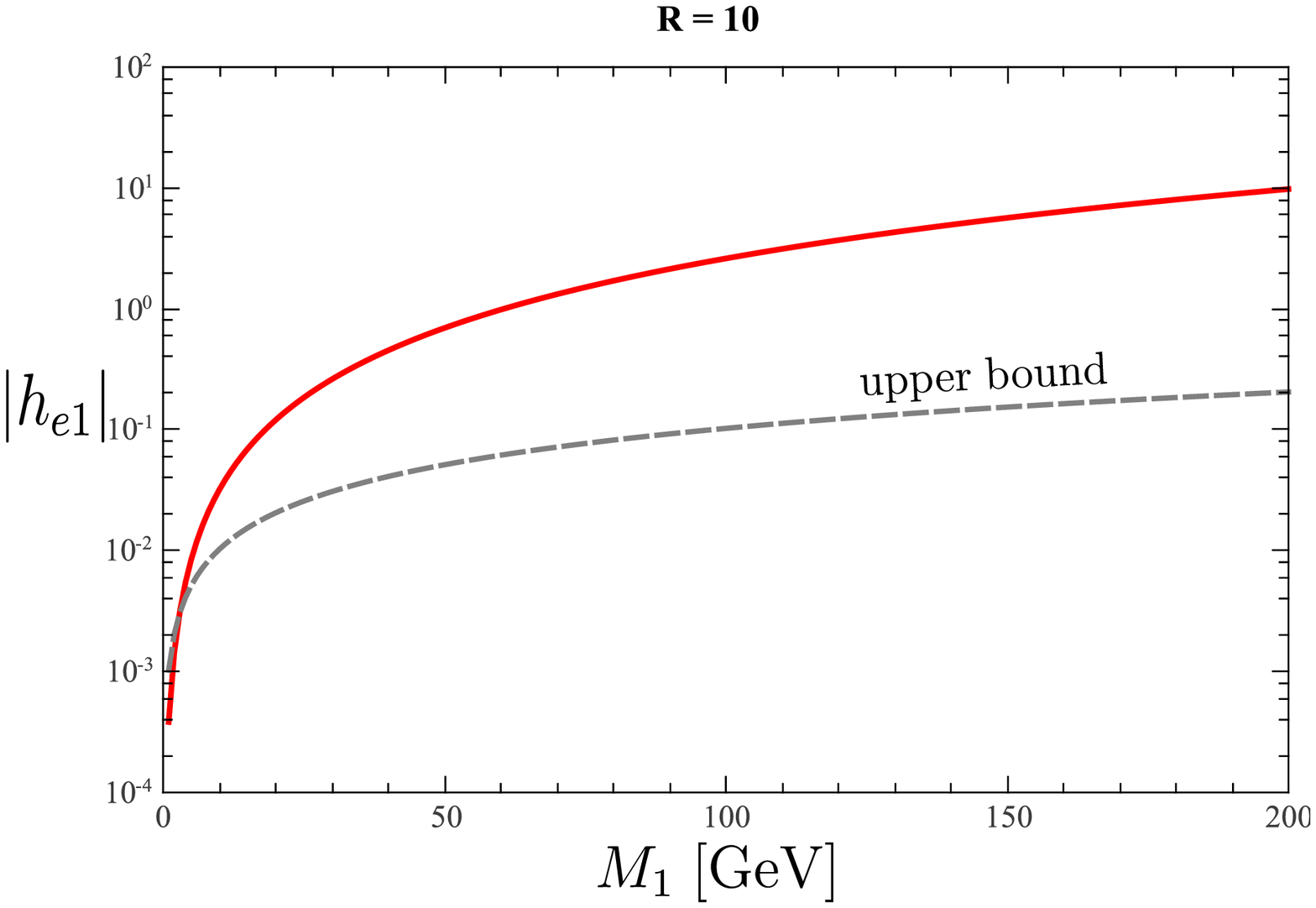}
\caption{
The fitting curve from CDM relic abundance fitting is presented by red solid lines.
Upper bound of $ |h_{e1}| $ given by the constraint 
Br($ \mu \rightarrow e \gamma $) $ < 2.4 \times 10^{-12} $
is presented by dash lines.
Different values of $ R $ = 2, 5, 10 are considered. 
} 
\label{flavour}
\end{center}
\end{figure}

As we mentioned in the introduction, this model is motivated by the dark matter and neutrino mass problems.
The existence of dark matter candidate is achieved by adding a $ Z_2 $ symmetry
and neutrino mass is completed through radiative effect contributed by additional particles $ N_i $ and $ \eta $.
However, these extra paticles not only provide the effective neutrino masses
, but also contribute to some other effects.
One significant effect is lepton flavour violation process.   
From this kind of process, we are able to get constraints on the parameter space of $h_{\alpha i}$. 
Due to the smallness of neutrino mass, the decay widths of lepton flavour violating processes
coming from neutrino masses are much smaller than present experiment limit, 
e.g. 
$ \mu \rightarrow e \gamma $, $ \mathrm{Br}(\mu \rightarrow e \gamma) < 10^{-40} $.  
However, the additional particles in this new model 
would contribute significantly to lepton flavour violating decay $ l_{\alpha} \rightarrow l_{\beta} \gamma $. 
Therefore, the experiment data would put constraint on the model by studying these processes.
These processes are showed by FIG. \ref{flavour_violation}.
The branch ratio of $ l_{\alpha} \rightarrow l_{\beta} \gamma $ is given by
\cite{Ma:2001mr}
\begin{align}
\mathrm{Br}(l_{\alpha} \rightarrow & l_{\beta} \gamma)
=
\dfrac{ \Gamma( l_{\alpha} \rightarrow l_{\beta} \gamma ) }
{ \Gamma( l_{\alpha} \rightarrow l_{\beta} \nu_{\alpha} \bar{\nu}_{\beta} ) } 
\notag
\\
&=
\dfrac{3 \alpha_{em}}{64 \pi G_F^2 m_0^4}
\left|
\sum_{i=1}^{3} h^{*}_{\alpha i} h_{\beta i} F \left( \dfrac{M_i^2}{m_0^2} \right)
\right|^2
\end{align}
where $ \alpha_{em} \equiv e^2 / 4 \pi $ is the electromagnetic fine structure constant, $ G_F $ is the Fermi constant
, and the function $ F(x) $ is defined by \cite{Ma:2001mr}
\begin{equation}
F(x) = \dfrac{1 - 6x + 3 x^2 + 2 x^3 - 6 x^2 \mathrm{ln} x}{6(1-x)^4}.
\label{Function}
\end{equation}
This function decreases as $ x $ increases.
As we mentioned in CDM abundance fitting, 
the mass spectrum $ M_1 < m_0 < M_2 $ is preferred to stay within validity of perturbative calculation.
So in this mass hierarchy, 
$ N_1 $ play the most significant role in these processes.
As a result, it gives constraints for the Yukawa couplings of $ N_1 $, $ h_{\alpha 1} $.
According to the latest experiment results, there are upper bounds for branch ratios
Br($ \tau \rightarrow \mu \gamma $) $ < 4.5 \times 10^{-8} $ \cite{Antipin:2007uh} 
and
Br($ \mu \rightarrow e \gamma $) $ < 2.4 \times 10^{-12} $ \cite{Adam:2011ch}.
Between these two bounds,
the process $ \mu \rightarrow e \gamma $ provides stronger constraint for our model.
Hence, we use this upper bound to cut the curve that we get from CDM relic abundance fitting.
FIG. \ref{flavour}
shows the result after using the upper bounds from decay process $ \mu \rightarrow e \gamma $.
As in the section of relic abundance, we consider different values of ratio $ R $.
For $ R = 2 $ case, the region of $ M_1 $ larger than 170 GeV is ruled out. 
For $ R = 5 $ case, only mass smaller than 20 GeV is allowed
For $ R = 10 $ case, the coupling constant is larger than 1 in some region, 
i.e. the non-perturbative Yukawa interaction regime.
Besides, almost all the parameter region is excluded by the requirement from $ \mu \rightarrow e \gamma $.
It implies that in CDM scenario $ m_0 $ cannot be too larger than $ M_1 $.
This result is consistent with the requirement of the theory being perturbative.
It also suggests the choice of the mass spectrum $ M_1 < m_0 < M_2 $. 
Although our result is based on the analysis of mixing matrix in Section II,
there would be no big difference because $ a_1 , a_2 $ are still of order one for such a highly mixed mass matrix.
By analyzing these plots, we found that
it also gives a constraint on $ m_0 $ that it must be not larger than 300 GeV. 
The well measured value of branch ratio provides a very strong constraint on this model.
Even in the region that is not excluded, the fitting curve is quite close to the upper limit. 
It means that if the measurement of $ \mu \rightarrow e \gamma $ can be improved,
it would be possible to detect the effect of flavour violation decay induced by $ N_i $ and $ \eta $.
On the other hand, 
if we still do not see any discrepancy beyond standard model in further experiment with better precisions, 
more region would has to be ruled out and endanger the validity of this model. 
\section{Muon anomalous magnetic moment}
The additional particles, $ N_i $ and $ \eta $, also contribute to the magnetic moment of leptons.
Similar with FIG. \ref{flavour_violation},
but the flavour of incoming and out-going leptons are now the same, i.e. $ \alpha = \beta $.
Among all kinds of leptons, the muon magnetic moment shows some interesting phenomenon. 
In 2001, E821 experiment at BNL found that there is discrepancy between the experimental value and prediction of muon anomaly $ a_{\mu} = (g-2)/2 $ from the Standard Model \cite{Brown:2001mga}.
It's possible to explain this deviation by new contribution. 
Thus, here we pay attention to the anomalous magnetic moment of muon. 
The muon anomalous magnetic moment contributed by additional particles is given by
\cite{Ma:2001mr}
\begin{equation}
\Delta a_{\mu} = \dfrac{m_{\mu}^2}{16\pi^2 m_0^2} 
\sum_{k} \left| h_{\mu k} \right|^2 F \left( \dfrac{M_k^2}{m_0^2} \right)
\end{equation}
where $ m_{\mu} $ is the mass of muon.
For muon anomaly, the discrepancy between experiment and theoretical prediction based on Standard Model is given by 
$ a_{\mu}^{\mathrm{EXP}} - a_{\mu}^{\mathrm{SM}} = (26.1 \pm 8.0) \times 10^{-10} $ \cite{Hagiwara:2011af}.
\begin{figure}[t]
\begin{center}
\includegraphics[width=0.8\linewidth]{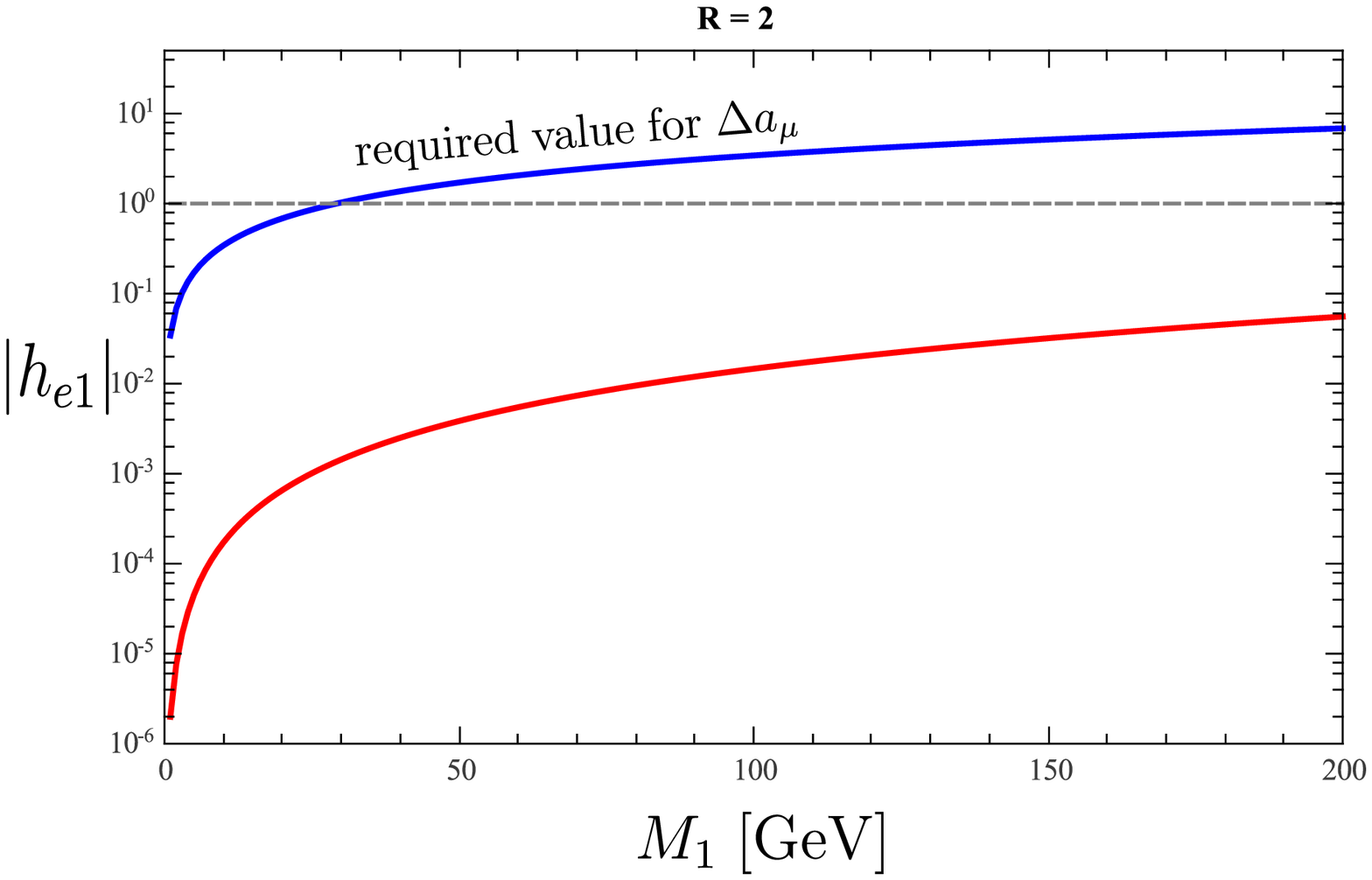}
\includegraphics[width=0.8\linewidth]{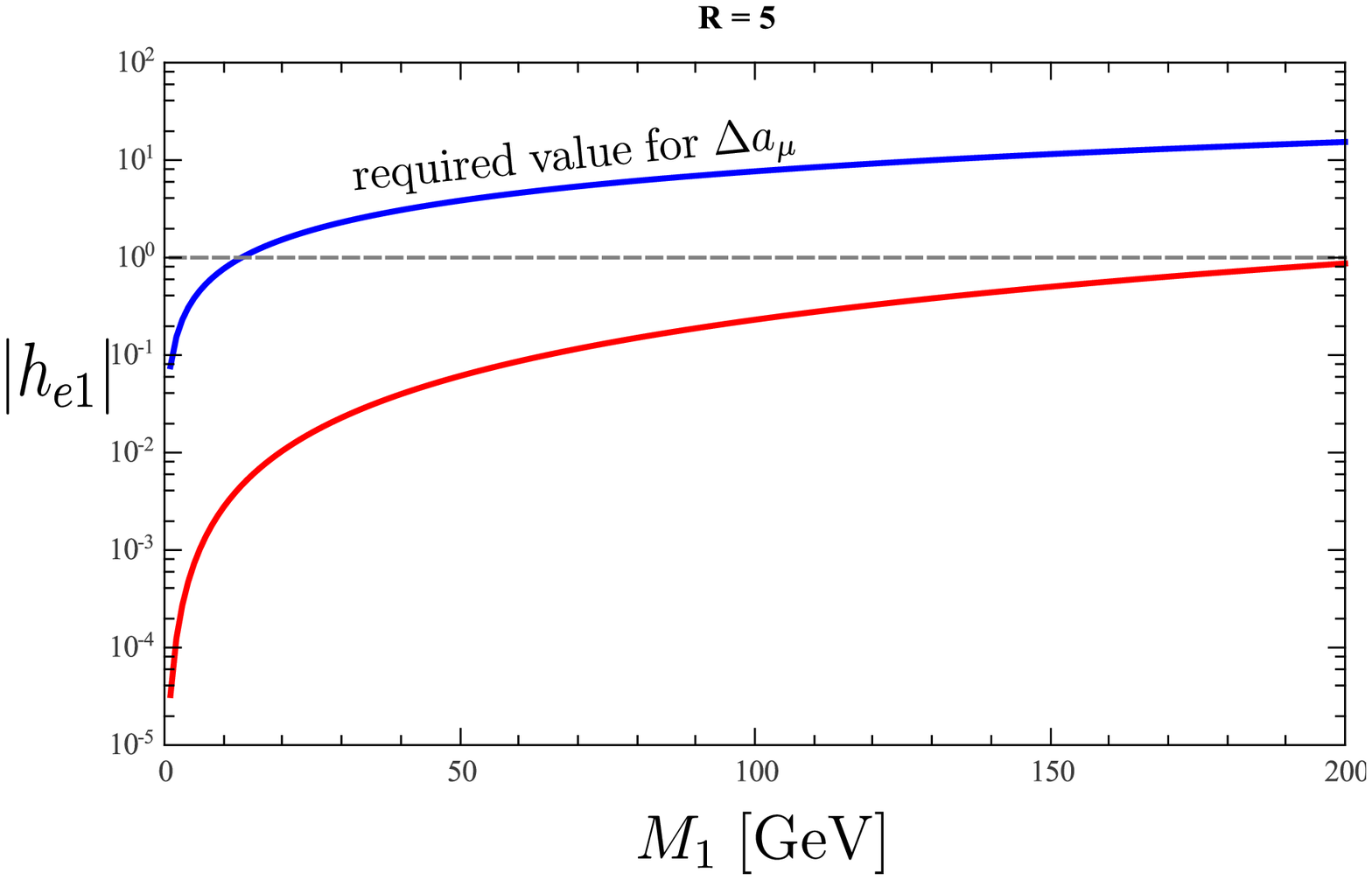}
\includegraphics[width=0.8\linewidth]{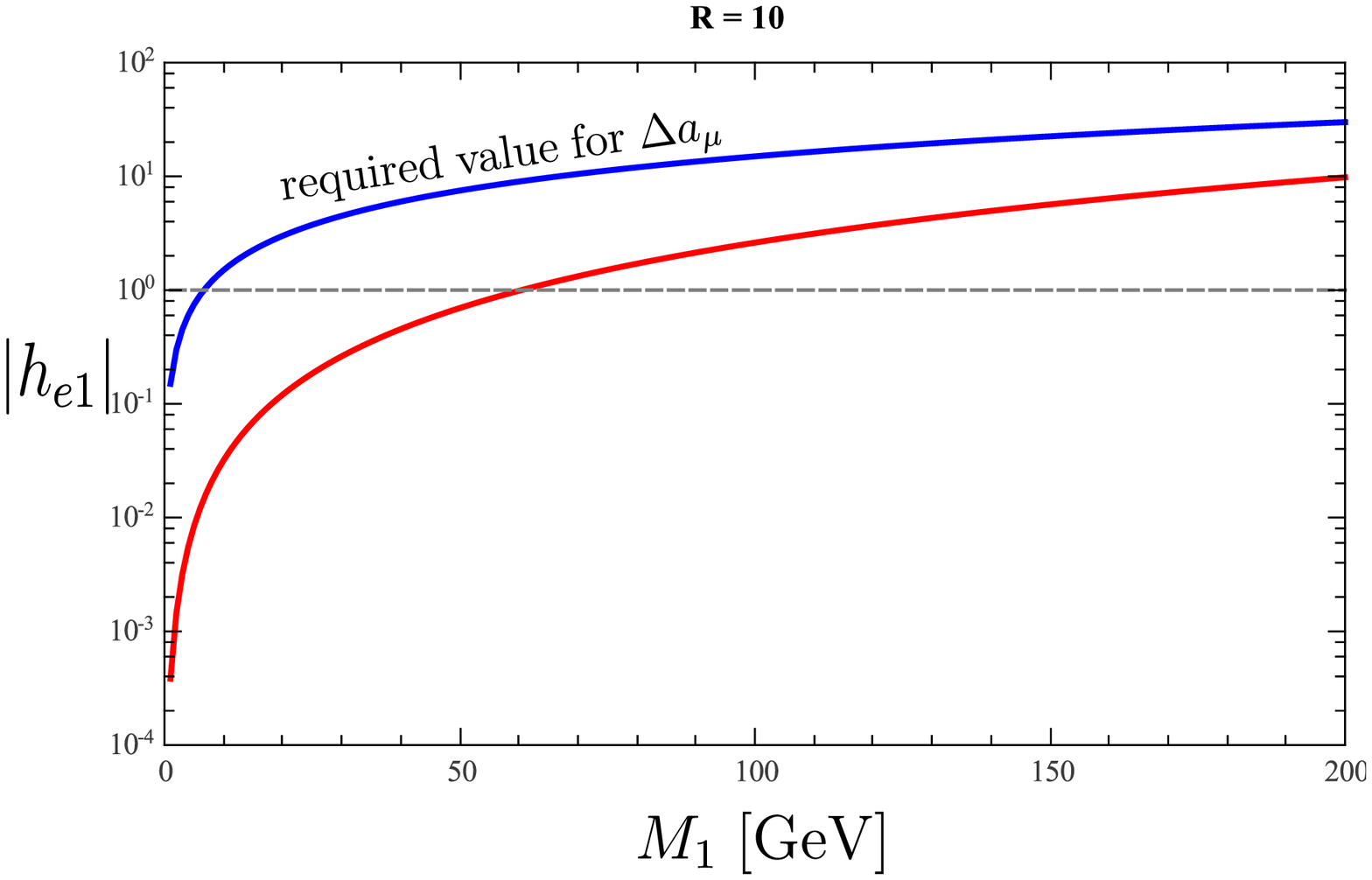}
\caption{
The fitting curves from relic density is presented by red lines.
And required values for explaining the observed muon anomaly is presented by blue lines. 
Different values of $ R = $ 2, 5, 10 are considered. 
}
\label{muon_anomaly}
\end{center}
\end{figure}
FIG. \ref{muon_anomaly} shows the fitting curve of relationship,$ |h_{e1}| $ to $ M_1 $, 
and the required value to account for the discrepancy of muon anomaly. 
From these figures, we can see that large coupling constants are needed for explaining muon anomaly. 
The values are very close to 1, i.e. the perturbative limit.
For $ R = 2 $ case, the coupling constant is larger than 1 for $ M_1 $ larger than 30 GeV. 
For $ R = 5 $ and $ R = 10 $, the coupling is below perturbative limit only while if $ M_1 $ is a few GeV.
The additional loop correction contributed by $ N_i $ and $ \eta $ cannot fully explain the observed muon anomaly. 

\section{Summary}
With the latest experimental data and most updated fit, 
we study Ma's radiative seesaw model with fewer assumptions and different scenarios.

In CDM scenario, 
we have done a more refined analysis than the previous work with the new information of $ \theta_{13} $ and mass mixing matrix .
Furthermore, 
if we require the theory to remain in the perturbative regime, 
the mass spectrum $ M_{1} <m_{0} < M_{2}$ is preferred.
By studying the flavour violation decay $ \mu \rightarrow e \gamma $,
a very strong constraint is obtained.
The ratio $ R \equiv m_0 / M_1 $ is favoured not to be larger than 10, 
and it show a upper bound roughly 300 GeV for $ m_0 $.
Compared to Ref. \cite{Suematsu:2009ww}, 
the constraint for $ M_1 $ is more precise, restricting it to be in the region below 200 GeV.
This result is consistent with the preferred mass spectrum $ M_1 < m_0 < M_2 $. 
In Ref. \cite{Sierra:2009}, it study a specific mass spectrum $ M_1 \ll M_2 < M_3 < m_0 $.
However, according to our result, it seems not viable in CDM scenario. 
We try to use additional contribution from new particles to explain observed muon anomaly.
Unfortunately, the muon anomalous magnetic moment induced by $ N_i $ and $ \eta $ is too small to fully account for the discrepancy.

In WDM scenario,
we use a new approach.
We show that WDM relic can be attained by thermal production with a subsequent entropy dilution process.
The entropy dilution can be accomplished by $ N_2 $ decay within the model, 
and no modification for this model is needed.
In this scenario, 
the mass spectrum $ M_1 < M_2 < m_0 $ is preferred for providing a required decay channel 
$ N_2 \rightarrow N_1 l \bar{l} $ via an intermediate virtual $ \eta $.
Different from previous works,
extremely small couplings are not required.

Direct detection of dark matter is also an important issue and has potential to give some restrictions for this model.
But, in this paper, we do not discuss the direct detection of dark matter.
Since $ N_i $ are Majorana fermions, 
effective interaction terms $ \bar{N_i} \gamma^{\mu} N_i $ and $ \bar{N_i} \sigma^{\mu \nu} N_i $ vanish.
Therefore, it's very hard to detect the elastic scattering between the target nuclei and dark matter $ N_i $.
One possibility is that the masses of $ N_1 $ and $ N_2 $ are highly degenerate.
In this case, probably we can detect inelastic scatterings between $ N_1 $ and $ N_2 $.
This case is discussed in Ref. \cite{Schmidt:2012yg}. 
Even in such a specific condition, 
the constraints given by direct detectors are still very limited.

For further study,
there are other potentially interesting phenomena including indirect signal from coannihilation and collider signature. 
For coannihilation events, data fitting is far more complicated due to the background analysis.
PAMELA collaboration reported their result indicating positron excess \cite{Adriani:2008zr}.
Ref. \cite{Suematsu:2010gv} try to explain the excess within this model.
But it needs an extremely large boost factor. 
For collider physics, some collider signatures of the model has been deduced in Ref. \cite{Sierra:2009,Aoki:2010tf}
As LHC runs and collects data, this kind of events can be examined in the future.

\bibliographystyle{apsrev4-1}

\end{document}